\documentclass[12pt]{iopart}
\renewcommand{\theequation}{\Roman{section}.\arabic{equation}}

\begin{document}

\title{Relativistic Wave Equations and Hydrogenic Atoms}
\author{B.A. Robson and S.H. Sutanto}
\address{Department of Theoretical Physics, Research School of Physical Sciences and Engineering, The Australian National University,  Canberra ACT 0200, Australia.}

\begin{abstract}
The transition probabilities for the components of both the Balmer
and Lyman $\alpha$-lines of hydrogenic atoms are calculated for
the nonrelativistic Schr\"odinger theory, the Dirac theory and the
recently developed eight-compo\-nent formalism. For large $Z$ it
is found that all three theories give significantly different
results.
\end{abstract}

%%%%%%%%%%%%%%%%%%%%%%%%%%%%%%%%%%%%%%%%%%
\section{Introduction}
\label{hydrogenic}
%%%%%%%%%%%%%%%%%%%%%%%%%%%%%%%%%%%%%%%%%%

Recently an eight-component  (8-C) relativistic wave equation for
spin-$\frac{1}{2}$ particles was proposed~\cite{RS1,S2} in an
attempt to place particles and antiparticles on a more symmetrical
basis than occurs in the Dirac equation. The 8-C equation gives
the same bound-state energy eigenvalue spectra for hydrogenic
atoms as the Dirac equation but the wavefunctions are different,
corresponding to a different Hamiltonian. This difference becomes
greater as the nuclear charge $Z$ increases. With a view to
ultimately distinguishing experimentally between the Dirac
equation and the 8-C equation, it is necessary to investigate
whether the different wavefunctions lead to different predictions
for observable quantities which depend explicitly upon the
wavefunctions, e.g. radiative transition probabilities between the
hydrogenic atomic bound states. Unfortunately, the 8-C equation
differs from the Dirac equation not only in having an enlarged
solution space (eight components versus four components), but also
in requiring the use of an indefinite inner product, which
complicates a direct comparison between the use of the two
relativistic wave equations.

In this paper, the relative transition probabilities for the
components of both the Balmer and Lyman $\alpha$-lines of
hydrogenic atoms will be discussed for the Schr\"odinger
(non-relativistic), Dirac and the 8-C wave equation formalisms.

%%%%%%%%%%%%%%%%%%%%%%%%%%%%%%%%%%%%%%%%%%%%%%%%%%%%%%%%%%%%%%%%%%%%%%%%
\section{Spontaneous emission and nonrelativistic Schr\"odi\-nger theory}
\label{emission-schrodinger}
%%%%%%%%%%%%%%%%%%%%%%%%%%%%%%%%%%%%%%%%%%%%%%%%%%%%%%%%%%%%%%%%%%%%%%%%

The normal decay of an excited atomic state takes place by the
spontaneous emission of radiation. This process can be described
within the framework of time-dependent perturbation theory and in
first order is given by Fermi's ``Golden Rule''~\cite{Friedrich90}
\begin{equation}
P_{fi} = \frac{2 \pi}{\hbar} \langle \phi_f|W|\phi_i \rangle^2 \,
\rho(E_f = E_i). \label{Golden-Rule}
\end{equation}
Here $P_{fi}$, is the total probability per unit time for
transitions from an initial state $\phi_i$ to all possible final
states $\phi_f$, $\rho (E_f)$ is the density of final states and
$W$ is the ``small perturbation'' causing the transition.

In the Schr\"odinger theory, the perturbing interaction (in the
Coulomb gauge) to first order is given by
\begin{equation}
W_S = - \frac{e}{2 \mu c} [ {\bf p} \cdot {\bf A} + {\bf A} \cdot
{\bf p}] \label{perturbing-interaction-1}
\end{equation}
where ${\bf p}$ is the momentum operator associated with the
electron, ${\bf A}$ is the vector potential operator associated
with electromagnetic field, $e = -|e|$ is the charge on the
electron and $\mu$ is the reduced mass of the hydrogenic atom.

The nonrelativistic Hamiltonian for the hydrogenic atom is
\begin{equation}
H_S = \frac{{\bf p}^2}{2 \mu} - \frac{Z e^2}{r}
\label{hamiltonian-0}
\end{equation}
where $r = |{\bf r}|$ (measured in atomic length units), ${\bf r}
= {\bf r}_e - {\bf r}_Z$, ${\bf r}_e$ and ${\bf r}_Z$ being the
spatial coordinates of the electron and the nucleus, respectively.

In the Schr\"odinger theory, spin may be included in the
two-component form and the eigenfunctions of the spin-independent
Hamiltonian (\ref{hamiltonian-0}) can be written
\begin{equation}
|njlm_j \rangle = \sum_{m, m_s} R_{nl} (r) Y_{lm} (\Omega)
\chi_{\frac{1}{2}m_s} C \left(l \mbox{$\frac{1}{2}$} j; m \, m_s
\, m_j \right). \label{eigenfunction-spin-independent}
\end{equation}
Here $n$, $j$, $l$ and $m$ are the usual principal, total angular
momentum, orbital angular and azimuthal quantum numbers,
respectively. The spin quantum number,  $m_s$, takes only the two
values $+\frac{1}{2}$ and $-\frac{1}{2}$ so that it is convenient
to represent the spin wavefunctions $\chi_{\frac{1}{2}m_s}$ in the
form
\begin{eqnarray}
\chi_{\frac{1}{2}\,\frac{1}{2}} = \left( \begin{array}{c} 1 \\ 0
\end{array} \right) & , & \chi_{\frac{1}{2} \, -\frac{1}{2}} =
\left( \begin{array}{c} 0 \\ 1 \end{array} \right) .
\label{spin-wave}
\end{eqnarray}
The coefficient $C \left(l \frac{1}{2} j; m \, m_s \, m_j \right)$
is the Clebsch-Gordan coefficient as defined by Rose~\cite{Rose57}
and which vanishes unless $m_j = m + m_s$. The radial wavefunction
$R_{nl} (r)$ is given by
\begin{eqnarray}
R_{nl} (r) & = & \frac{1}{(2 l + 1)!} \left[ \frac{(n+l)!}{(n - l
- 1)! 2n} \right]^{\frac{1}{2}} \left( \frac{2Z}{n}
\right)^{\frac{1}{2}} \left(\frac{2Zr}{n} \right)^l \nonumber \\
 &   & \times \, \exp (-Zr/n) F(l+1-n,2l+2;2Zr/n)
\label{radial-wave}
\end{eqnarray}
where
\begin{equation}
F(a,b;x) = 1 + \frac{ax}{b} + \frac{a(a+1)x}{b(b+1)2!} + \, ...
\label{confluent-hypergeometric}
\end{equation}
is the confluent hypergeometric function. The eigenenergies are
given by the Bohr terms
\begin{equation}
E_n = - \frac{\mu e^4}{2 \hbar^2} \frac{Z^2}{n^2} \: , \: n =
1,2,3, ... \label{eigenenergy}
\end{equation}

From (\ref{Golden-Rule}) and (\ref{perturbing-interaction-1}) the
probability per unit time for an atomic transition from an initial
state $| \phi_i \rangle \equiv |n j l m_j \rangle$ to a final
state $| \phi_f \rangle \equiv |n^\prime j^\prime l^\prime
m_{j^\prime} \rangle$ accompanied by the emission of a photon with
wave vector ${\bf k}_\lambda$, angular frequency $\omega_\lambda =
c |{\bf k}_\lambda |$, and polarization vector $\mbox{\boldmath
$\pi$}_\lambda$ of unit length is given by
\begin{equation}
P_{f i} = \frac{1}{2 \pi \hbar} \frac{e^2 \omega_\lambda}{\mu^2
c^3} \left| \langle n^\prime j^\prime l^\prime m_{j^\prime} | e^{-
i {\bf k}_\lambda \cdot {\bf r}} \mbox{\boldmath $ \pi$}_\lambda
\cdot {\bf p} |n j l m_j \rangle \right|^2 .
\label{transition-prob-1}
\end{equation}
Thus the total probability per unit time for an atomic transition
from all the initial states with the same $n$, $j$, and $l$ to all
the final states with the same $n^\prime$, $j^\prime$ and
$l^\prime$ accompanied by the emission of a photon of arbitrary
polarization in any direction is
\begin{eqnarray}
P_T & = & \sum_{m_{j^\prime}, m_j} \sum_{\lambda} \, \int P_{fi}
\, d\Omega_k \nonumber \\
 &  =  & \frac{2 e^2}{\mu^2 c^2 \hbar} \sum_{m_{j^\prime}, m_j} \sum_{\lambda} k_{\lambda} \left| \langle n^\prime j^\prime l^\prime m_{j^\prime} | e^{- i {\bf k}_\lambda \cdot {\bf r}} \mbox{\boldmath $\pi$}_\lambda \cdot {\bf p} | n j l m_j \rangle \right|^2
\label{PT-formula}
\end{eqnarray}

Taking the $z$-axis along ${\bf k}_\lambda$, i.e. ${\bf k}_\lambda
= k_\lambda \hat{{\bf e}}_z$, the two polarization components
$\mbox{\boldmath  $\pi$}_\lambda$ can be represented by
$\mbox{\boldmath $\pi$}_\pm = \mp \frac{1}{\sqrt{2}} (\hat{{\bf
e}}_x \pm i \hat{{\bf e}}_y)$. Using the plane wave expansion in
terms of spherical harmonics and spherical Bessel functions
\begin{equation}
e^{-i {\bf k}_\lambda \cdot {\bf r}} = \sum_{L} [4 \pi (2 L +
1)]^{1/2} i^{-L} Y_{L0} (\Omega) j_L (k_\lambda r)
\label{harmonic-bessel}
\end{equation}
we obtain
\begin{eqnarray}
\lefteqn{\langle n^\prime j^\prime l^\prime m_{j^\prime} | e^{- i
{\bf k}_\lambda \cdot {\bf r}} \mbox{\boldmath $\pi$}_{\pm} \cdot
{\bf p} | n j l m_j \rangle = } \nonumber \\
 &   & i \sum_{L} (2L + 1) i^{-L} \sum_{m} C(l^\prime \mbox{$\frac{1}{2}$} j^\prime ; m \pm 1 m_j - m m_{j^\prime}) C(l \mbox{$\frac{1}{2}$} j ; m m_j-m m_j) \nonumber \\
 & \times & \left[ \langle n^\prime l^\prime \parallel j_L (k_\lambda r) \parallel F_{nl}^{(+)} \rangle C(l^\prime\, L\, l+1\, ; 0\, 0\, 0) C(l+1\, 1\, l ; 0\, 0\, 0) \right. \nonumber \\
 &    & \left. \times C(l+1\, L\, l^\prime\, ; m \pm 1\, 0\, m \pm 1) C(l\, 1\, l+1; m, \, \pm 1\, m \pm 1) \right. \nonumber \\
 &    &  \left. +  \langle n^\prime l^\prime \parallel j_L (k_\lambda r) \parallel F_{nl}^{(-)} \rangle C(l^\prime\, L\, l-1 ; 0\, 0\, 0) C(l-1\, 1\, l ; 0\, 0\, 0) \right. \nonumber \\
 &    & \left. \times C(l-1\, L\, l^\prime ; m \pm 1\, 0\, m \pm 1) C(l\, 1\, l-1; m,\, \pm 1\, m \pm 1) \right] .
\label{ptschr}
\end{eqnarray}
Here the functions $ |F_{nl}^{(\pm)} \rangle $ are given by
\begin{eqnarray}
F_{nl}^{(+)} (r) = \left( \frac{d}{dr} - \frac{l}{r} \right)
R_{nl}(r) & and & F_{nl}^{(-)} (r) = \left( \frac{d}{dr} -
\frac{l+1}{r} \right) R_{nl}(r) \label{fplusminus}
\end{eqnarray}
and we have used (A38) of Bethe and Salpeter~\cite{BS77}. The
quantities $ \langle n^{\prime} l^{\prime} \parallel
j_L(k_{\lambda}r) \parallel F_{nl}^{(\pm)}\rangle $ are radial
reduced matrix elements. Using (~\ref{ptschr}) in
(~\ref{PT-formula}) gives the transition probabilities for the
components of the Balmer and Lyman $\alpha$-lines for various
hydrogenic atoms presented in the columns labelled S in
Tables~\ref{allow-1} - \ref{forbid-92}.

%%%%%%%%%%%%%%%%%%%%%%%%%%%%%%%%%%%%%%%%
\section{Dirac theory}
\label{Dirac-Theory}
%%%%%%%%%%%%%%%%%%%%%%%%%%%%%%%%%%%%%%%%

In the Dirac theory of spontaneous emission, the perturbing
interaction (in the Coulomb gauge) to first order is
\begin{equation}
W_D = - e \mbox{\boldmath $\alpha$} \cdot {\bf A}
\label{1-order-interaction}
\end{equation}
where
\begin{equation}
\mbox{\boldmath $\alpha$} = \left( \begin{array}{cc}
                {\bf 0} & \mbox{\boldmath $\sigma$} \\
                \mbox{\boldmath $\sigma$} & {\bf 0}
               \end{array} \right)
\label{alpha-matrix}
\end{equation}
$\mbox{\boldmath $\sigma$}$ being the usual Pauli spin vector with
components
\begin{equation}
\sigma_x = \left( \begin{array}{cc}
            0 & 1 \\
            1 & 0
          \end{array}  \right) \, , \,
\sigma_y = \left( \begin{array}{cc}
            0 & -i \\
            i & 0
          \end{array}  \right) \, , \,
\sigma_z = \left( \begin{array}{cc}
            1 & 0 \\
            0 & -1
           \end{array} \right).
\label{Pauli-matrix}
\end{equation}

The Dirac Hamiltonian for the hydrogenic atom
\begin{equation}
H_D = c \mbox{\boldmath $\alpha$} \cdot {\bf p} + \beta \mu c^2 -
\frac{Z e^2}{r} \label{Hamiltonian-hydrogenic}
\end{equation}
with
\begin{equation}
\beta = \left( \begin{array}{cc}
        {\bf 1_2} & {\bf 0} \\
        {\bf 0} & {\bf -1_2}
           \end{array} \right).
\label{beta-matrix}
\end{equation}
The eigenfunctions of the Hamiltonian
(\ref{Hamiltonian-hydrogenic}) can be writen
\begin{equation}
|njlm_j \rangle = \sum_{m, m_s} \left[ \begin{array}{c}
                g (r) C( l \frac{1}{2}j; m m_s m_j) Y_{l m} (\Omega) \chi_{\frac{1}{2} m_s} \\
                i f(r) C( \bar{l} \frac{1}{2}j; m m_s m_j) Y_{\bar{l} m} (\Omega) \chi_{\frac{1}{2} m_s}
                       \end{array} \right]
\label{eigenfunctions-Hamiltonian-hydrogenic}
\end{equation}
where $\bar{l} = l \pm 1$ for $j = l \pm \frac{1}{2}$ and the
radial functions are~\cite{BS77}
\begin{eqnarray}
g (r) & = & - \frac{[\Gamma (2 \gamma + \tilde{n} +
1]^{\frac{1}{2}}]}{\Gamma (2 \gamma +
1)(\tilde{n}!)^{\frac{1}{2}}} \, \left[ \frac{(1 + \epsilon)}{4 N
(N - \kappa)} \right]^{\frac{1}{2}} \, \left(\frac{2Z}{N}
\right)^{\frac{3}{2}} \, \left(\frac{2 Z r}{N} \right)^{\gamma -
1} \nonumber \\
 &   & \times \exp(-Zr/N) [-\tilde{n} F(1 - \tilde{n}, 2\gamma + 1; 2Zr/N) \nonumber \\
 &   & + (N - \kappa) F(-\tilde{n}, 2 \gamma+1; 2Zr/N)]
\label{g-function}
\end{eqnarray}
and
\begin{eqnarray}
f (r) & = & - \frac{[\Gamma (2 \gamma + \tilde{n} +
1]^{\frac{1}{2}}]}{\Gamma (2 \gamma +
1)(\tilde{n}!)^{\frac{1}{2}}} \, \left[ \frac{(1 - \epsilon)}{4 N
(N - \kappa)} \right]^{\frac{1}{2}} \, \left(\frac{2Z}{N}
\right)^{\frac{3}{2}} \, \left(\frac{2 Z r}{N} \right)^{\gamma -
1} \nonumber \\
 &   & \times \exp(-Zr/N) [\tilde{n} F(1 - \tilde{n}, 2\gamma + 1; 2Zr/N) \nonumber \\
 &   & + (N - \kappa) F(-\tilde{n}, 2 \gamma+1; 2Zr/N)].
\label{f-function}
\end{eqnarray}
Here
\begin{equation}
\begin{array}{ccccc}
\kappa & = & -(l + 1) &   &  \mbox{for} \, j = l + \frac{1}{2}\\
       & = & + l      &   &  \mbox{for} \, j = l - \frac{1}{2}
\end{array}
\label{kappa}
\end{equation}
\begin{equation}
\gamma = \left[ \kappa^2 - \alpha^2 Z^2 \right]^{\frac{1}{2}}
\hspace{8mm} \mbox{with} \, \alpha = e^2/\hbar c. \label{gamma}
\end{equation}
\begin{equation}
\tilde{n} = n - |\kappa| \label{n-tilde}
\end{equation}
\begin{equation}
\epsilon = \left[ 1 + \frac{\alpha^2 Z^2}{(\tilde{n} + \gamma)^2}
\right]^{-\frac{1}{2}} \label{epsilon}
\end{equation}
and
\begin{equation}
N = \left[ n^2 - 2 \tilde{n} (|\kappa| - \gamma)
\right]^{\frac{1}{2}}. \label{N}
\end{equation}

Corresponding to (\ref{transition-prob-1}) we have
\begin{equation}
P_{f i} = \frac{1}{2 \pi \hbar} \frac{e^2 \omega_\lambda}{c}
\left| \langle n^\prime j^\prime l^\prime m_{j^\prime} | e^{- i
{\bf k}_\lambda \cdot {\bf r}}\mbox{\boldmath $\pi$}_{\lambda}
\cdot \mbox{\boldmath $\alpha$} | n j l m_j \rangle \right|^2
\label{trans-prob dirac}
\end{equation}
Using spherical components of $ \mbox{\boldmath $\alpha$}$ :
\begin{eqnarray}
\alpha_0 = \alpha_z & , & \alpha_{\pm} = \mp \frac{1}{\sqrt{2}}
(\alpha_x \pm i \alpha_y) \label{alpha plusminus}
\end{eqnarray}
and the planewave expansion (\ref{harmonic-bessel}) one obtains
\begin{eqnarray}
\lefteqn{ \langle n^\prime j^\prime l^\prime m_{j^\prime} | e^{- i
{\bf k}_\lambda \cdot {\bf r}} \alpha_{\pm} | n j l m_j \rangle =}
\nonumber \\ &   & \mp i \sqrt{2} \sum_{L} i^{-L} (2L+1) \left[
\langle g^{\prime} \parallel j_L(k_{\lambda} r)
\parallel f \rangle C(l^{\prime}\, \mbox{$\frac{1}{2}$}\,
j^{\prime} ; m_j \pm \mbox{$\frac{1}{2}$},\,\pm \mbox{$
\frac{1}{2}$}\, m_j \pm 1)\right. \nonumber \\ &   & \left. \times
C(\bar{l}^{\prime}\, \mbox{$\frac{1}{2}$}\, j^{\prime} ; m_j \pm
\mbox{$ \frac{1}{2}$},\, \mp \mbox{$ \frac{1}{2}$}\, m_j)
C(\bar{l}^{\prime}\, L\, l^{\prime} ; m_j \pm \mbox{$
\frac{1}{2}$}\, 0\, m_j\pm \mbox{$ \frac{1}{2}$})
C(l^{\prime}\,L\,\bar{l} ;0\,0\,0) \right. \nonumber \\ &   &
\left. - \langle f^{\prime} \parallel j_L(k_{\lambda} r)
\parallel g \rangle   C(\bar{l}^{\prime}\, \mbox{$\frac{1}{2}$}\,
j^{\prime} ; m_j \pm \mbox{$ \frac{1}{2}$},\,\pm \mbox{$
\frac{1}{2}$}\, m_j \pm 1)  C(l\, \mbox{$ \frac{1}{2}$}\, j ; m_j
\pm \mbox{$\frac{1}{2}$},\,\mp \mbox{$\frac{1}{2}$}\, m_j) \right.
\nonumber \\ &   & \left. \times C(l\, L\, \bar{l}^{\prime} ;
m_j\mbox{$ \pm \frac{1}{2}$}\, 0\, m_j \pm \mbox{$ \frac{1}{2}$})
C(\bar{l}^{\prime}\, L\, l ; 0\,0\,0) \right] \label{PT-dirac}
\end{eqnarray}
where $ \alpha_{\pm} \equiv \mbox{\boldmath $\pi$}_{\pm} \cdot
\mbox{\boldmath $\alpha$}$. Using (\ref{PT-dirac}) in the relation
corresponding to (\ref{PT-formula}) :
\begin{equation}
P_{T} = \frac{2e^2}{\hbar} \sum_{m_{j^\prime},m_j} \sum_{\lambda}
k_{\lambda}\left| \langle n^\prime j^\prime l^\prime m_{j^\prime}
| e^{- i {\bf k}_\lambda \cdot {\bf r}} \mbox{\boldmath
$\pi$}_\lambda \cdot \mbox{\boldmath $\alpha$} | n j l m_j \rangle
\right|^2 \label{PT-diracfinal}
\end{equation}
gives the transition probabilities for the components of the
Balmer and Lyman $\alpha$-lines for various hydrogenic atoms
presented in the columns labelled D in Tables~\ref{allow-1} -
\ref{forbid-92}.

%%%%%%%%%%%%%%%%%%%%%%%%%%%%%%%%%%%%%%%%%%%%%%%%%%%%%%
\section{Eight-component theory}
\label{eight-component}
%%%%%%%%%%%%%%%%%%%%%%%%%%%%%%%%%%%%%%%%%%%%%%%%%%%%%%

The eight-component equation for hydrogenic atom in the presence
of an external electromagnetic field is, in the Weyl
representation, given~\cite{RS1} by
\begin{equation}
\left(i \hbar \frac{\partial}{\partial t} \mbox{\boldmath{$1_8$}}
\right) \Psi_{FV1/2} = H_{FV1/2} \Psi_{FV1/2}  \label{8-component}
\end{equation}
where
\begin{equation}
H_{FV1/2} = \left(\begin{array}{cc} H_\xi & 0 \\ 0 & H_\eta
\end{array} \right)  \label{8-C-hamiltonian}
\end{equation}
and
\begin{eqnarray}
H_\xi & = &(\tau_3 + i \tau_2) \otimes \left(\frac{\hbar^2}{2 \mu}
[- {\bf D}^2 {\bf 1_2} + \frac{i e}{\hbar c}
\mbox{\boldmath{$\sigma$}} \cdot ({\bf E} + i {\bf B})] \right)
\label{Hamiltonian-xi} \nonumber \\
 &   & + \tau_3 \otimes (\mu c^2 \mbox{\boldmath{$1_2$}}) + e A_0
 \mbox{\boldmath{$1_4$}}, \\
H_\eta & = &(\tau_3 + i \tau_2) \otimes \left(\frac{\hbar^2}{2
\mu} [- {\bf D}^2 {\bf 1_2} - \frac{i e}{\hbar c}
\mbox{\boldmath{$\sigma$}} \cdot ({\bf E} - i {\bf B})] \right)
\nonumber \\
 &   & + \tau_3 \otimes (\mu c^2 \mbox{\boldmath{$1_2$}}) + e A_0
 \mbox{\boldmath{$1_4$}} \label{Hamiltonian-eta}
\end{eqnarray}
where $\tau_i$ are the standard Pauli matrices, $\otimes$ is the
Kronecker (direct) product, {\bf E} and {\bf B} are the
electromagnetic field intensities and ${\bf D} = {\bf \partial} +
(ie/\hbar c) {\bf A}$ is the usual minimal coupling. Thus in the
absence of an external electromagnetic field, the eight-component
Hamiltonian for a hydrogenic atom is (setting $A_0 = -Ze/r$)
\begin{equation}
H_8 = \frac{{\bf p}^2}{2 \mu} X -\frac{iZe^2 \hbar}{2 \mu c r^3}
\mbox{\boldmath{$\Sigma \cdot r$}} + \mu c^2 Y - \frac{Z e^2}{r}
{\bf 1_8}. \label{free-eight-hamiltonian}
\end{equation}
where $X = [{\bf 1_2} \otimes (\tau_3 + i \tau_2) \otimes {\bf
1_2}]$, ${\bf \Sigma} = [\tau_3 \otimes (\tau_3 + i \tau_2)
\otimes {\bf \sigma}]$ and $Y = [{\bf 1_2} \otimes \tau_3 \otimes
{\bf 1_2}]$.

It should be noted that in the above we have assumed the
decoupled form of the eight-component theory so that the inner
product is given~\cite{S2} by
\begin{equation}
\langle \Psi | \Psi \rangle = \int \Psi^{\dagger} (x) \tau_5 \Psi
(x) \, d^3x   \label{indefinite-inner-product}
\end{equation}
where $\tau_5 = \tau_1 \otimes \tau_3 \otimes {\bf 1}_2$.

Choosing the Coulomb gauge, {\it i.e.} ${\bf E} = -(1/c) \partial
{\bf A}/ \partial t$, ${\bf B} = \nabla \times {\bf A}$ for the
external field in (\ref{Hamiltonian-xi}) and
(\ref{Hamiltonian-eta}) gives the perturbing interaction to the
first order for spontaneous emission as the sum of three terms :
\begin{equation}
W_8 = W_8^{(1)} + W_8^{(2)} + W_8^{(3)} \label{W-8}
\end{equation}
where
\begin{eqnarray}
W_8^{(1)} & = & - \frac{e}{2 \mu c} [ {\bf p} \cdot {\bf A} + {\bf
A} \cdot {\bf p}] X    \label{W-8-1} \\ W_8^{(2)} & = & -\frac{i
e\hbar}{2 \mu c^2} \left[ \tau_3 \otimes (\tau_3 + i \tau_2)
\otimes \mbox {\boldmath $\sigma$} \cdot \frac{\partial}{\partial
t} {\bf A} \right] \label{W-8-2} \\ W_8^{(3)} & = & - \frac{e
\hbar}{2 \mu c} \left[ {\bf 1_2} \otimes (\tau_3 + i \tau_2)
\otimes \mbox{\boldmath $\sigma$} \cdot \mbox{\boldmath $\nabla$}
\times {\bf A} \right]   \label{W-8-3}
\end{eqnarray}
The eigenfunctions of the Hamiltonian
(\ref{free-eight-hamiltonian}) $|njlm_j \rangle$ can be written
as~\cite{RS1}
\begin{equation}
\sum_{m, m_s} \left[ \begin{array}{c} \bar{g} (r) \left\{ C(l
\frac{1}{2} j;m m_s m_j) Y_{l m} (\Omega) + i \bar{\kappa}
C(\bar{l} \frac{1}{2} j; m m_s m_j) Y_{\bar{l} m} (\Omega)
\right\} \chi_{\frac{1}{2} m_s} \\ \bar{f} (r)\left\{ C(l
\frac{1}{2} j;m m_s m_j) Y_{l m} (\Omega) + i \bar{\kappa}
C(\bar{l} \frac{1}{2} j; m m_s m_j) Y_{\bar{l} m} (\Omega)
\right\} \chi_{\frac{1}{2} m_s} \\ \bar{g} (r)\left\{ C(l
\frac{1}{2} j;m m_s m_j) Y_{l m} (\Omega) - i \bar{\kappa}
C(\bar{l} \frac{1}{2} j; m m_s m_j) Y_{\bar{l} m} (\Omega)
\right\} \chi_{\frac{1}{2} m_s}
\\ \bar{f} (r)\left\{ C(l \frac{1}{2} j;m m_s m_j) Y_{l m}
(\Omega) - i \bar{\kappa} C(\bar{l} \frac{1}{2} j; m m_s m_j)
Y_{\bar{l} m} (\Omega) \right\} \chi_{\frac{1}{2} m_s}
\end{array} \right]
\label{8-component-eigenfunction}
\end{equation}
where $\bar{l} = l \pm 1$ for $j =  l \pm \frac{1}{2}$ and the
radial functions are given by
\begin{eqnarray}
\bar{g} (r) & = & \frac{|\Lambda|^2 [ \Gamma(2 \bar{\gamma} +
\tilde{n}^{\prime} +1)]^{\frac{1}{2}}}{\Gamma(2 \bar{\gamma} +
2)[Z(\tilde{n}^{\prime} - 1)!]^{\frac{1}{2}}} \left\{ 2 |\Lambda|
r \right\}^{\bar{\gamma}} \left\{1 + \epsilon + \alpha^2 Z/r
\right\} \nonumber \\
 &   & \times \frac{\exp(-|\Lambda| r)}{\{2 (1 - \bar{\kappa}^2)\}^{\frac{1}{2}}} F(1 - \tilde{n}^{\prime}, 2 \bar{\gamma} + 2; 2 |\Lambda| r) \label{g-function-2} \\
\bar{f} (r) & = & \frac{|\Lambda|^2 [ \Gamma(2 \bar{\gamma} +
\tilde{n}^{\prime} +1)]^{\frac{1}{2}}}{\Gamma(2 \bar{\gamma} +
2)[Z(\tilde{n}^{\prime} - 1)!]^{\frac{1}{2}}} \left\{ 2 |\Lambda|
r \right\}^{\bar{\gamma}} \left\{1 - \epsilon - \alpha^2 Z/r
\right\} \nonumber \\
 &   & \times \frac{\exp(-|\Lambda| r)}{\{2 (1 - \bar{\kappa}^2)\}^{\frac{1}{2}}} F(1 - \tilde{n}^{\prime}, 2 \bar{\gamma} + 2; 2 |\Lambda| r)   \label{f-function-2}
\end{eqnarray}
Here
\begin{equation}
\begin{array}{ccccc}
\bar{\gamma} & = & \gamma - 1 &    &  \mbox{for} \: j = l +
\frac{1}{2} \\
 & = & \gamma &   & \mbox{for} \: j = l - \frac{1}{2}
\end{array}
\label{gamma-bar}
\end{equation}
\begin{equation}
\begin{array}{ccccc}
Z \alpha \bar{\kappa} & = & \kappa \pm \gamma &   & \mbox{for} \:
j = l \pm \frac{1}{2}
\end{array}
\label{Z-alpha-kappa}
\end{equation}
\begin{equation}
\begin{array}{ccccc}
\tilde{n}^{\prime} + \bar{\gamma} = \tilde{n} + \gamma &   &
\end{array}
\label{n-tilde-gamma}
\end{equation}
and
\begin{equation}
|\Lambda| = Z / N. \label{Lambda}
\end{equation}

In the eight-component theory, corresponding to
(\ref{transition-prob-1}) we have
\begin{equation}
P_{f i} = \frac{1}{2 \pi \hbar} \frac{e^2 k_\lambda}{\mu^2 c^2}
\left| M_1^{\lambda} + M_2^{\lambda} + M_3^{\lambda} \right|^2
\label{transition-prob-8C}
\end{equation}
where
\begin{eqnarray}
M_1^{\lambda} & = & \langle n^\prime j^\prime l^\prime
m_{j^\prime} | \tau_5 e^{- i {\bf k}_{\lambda} \cdot {\bf
r}}\mbox{\boldmath$\pi$}_\lambda \cdot {\bf p} X | n j l m_j
\rangle \label{M-1-8C}
\\ M_2^{\lambda} & = & - \frac{\hbar k_\lambda}{2} \langle n^\prime
j^\prime l^\prime m_{j^\prime} | \tau_5 e^{- i {\bf k}_{\lambda}
\cdot {\bf r}}\mbox{\boldmath $\pi$}_\lambda \cdot {\bf \Sigma}| n
j l m_j \rangle \label{M-2-8C} \\ M_3^{\lambda} & = & -
\frac{\hbar k_\lambda}{2} \langle n^\prime j^\prime l^\prime
m_{j^\prime} | \tau_5 e^{- i {\bf k}_{\lambda} \cdot {\bf r}}
\mbox{\boldmath $\pi$}_\lambda \cdot {\bf \Sigma}^\prime | n j l
m_j \rangle \label{M-3-8C}
\end{eqnarray}
are the matrix elements corresponding to $W_8^{(1)}$, $W_8^{(2)}$
and $W_8^{(3)}$, respectively, and
\begin{equation}
{\bf  \Sigma}^\prime = {\bf 1}_2 \otimes (\tau_3 + i \tau_2)
\otimes \mbox{\boldmath$\sigma$}.
\end{equation}
Using the plane wave expansion (\ref{harmonic-bessel}) and the
polarization components $\mbox{\boldmath $\pi$}_{\pm}$, one
obtains
\begin{eqnarray}
M_1^{(\pm)} & = & 2 i \sum_{L} (2L +1) i^{-L} \sum_{m_s}
\left[C(l^\prime \mbox{$\frac{1}{2}$} j^\prime ; m_j-m_s \pm 1 m_s
m_j \pm 1) \right. \nonumber \\
 &   & \left. \times C(l \mbox{$\frac{1}{2}$} j ; m_j-m_s m_s m_j) \right.
\nonumber \\
 &  & \left. \times \left\{ \langle \bar{g}^\prime +
 \bar{f}^\prime \parallel j_L (k_\lambda r) \parallel
 F_{njl}^{(+)} \rangle C(l+1 L l^\prime ; m_j-m_s \pm 1 0 m_j-m_s
 \pm 1) \right. \right. \nonumber \\
 &  & \left. \left. \times C(l^\prime L l+1; 0 0 0) C(l 1 l+1; m_j-m_s, \pm 1 m_j-m_s \pm 1) \right. \right. \nonumber \\ &  & \left. \left. \times C(l+1 1 l; 0
 0 0) \right. \right. \nonumber \\
 &  & \left. \left. + \langle \bar{g}^\prime + \bar{f}^\prime \parallel j_L (k_\lambda r) \parallel
 F_{njl}^{(-)} \rangle C(l-1 L l^\prime ; m_j-m_s \pm 1 0 m_j-m_s
 \pm 1) \right. \right. \nonumber \\
 &  & \left. \left. \times C(l^\prime L l-1; 0 0 0) C(l 1 l-1; m_j-m_s, \pm 1 m_j-m_s \pm 1) \right. \right. \nonumber \\ &  & \left. \left. \times C(l-1 1 l; 0
 0 0) \right\} \right. \nonumber \\
 &  & \left. - \bar{\kappa}^\prime \bar{\kappa} C(\bar{l}^\prime
 \mbox{$\frac{1}{2}$} j^\prime ; m_j-m_s \pm 1 m_s m_j \pm 1)
 C(\bar{l} \mbox{$\frac{1}{2}$} j ; m_j-m_s m_s m_j) \right.
 \nonumber \\
 &  & \left. \times \left\{ \langle \bar{g}^\prime +
 \bar{f}^\prime \parallel j_L (k_\lambda r) \parallel
 F_{njl}^{(+)} \rangle C(\bar{l}+1 L \bar{l}^\prime ; m_j-m_s \pm 1 0 m_j-m_s
 \pm 1)  \right. \right. \nonumber \\
 &  & \left. \left. \times C(\bar{l}^\prime L \bar{l}+1; 0 0 0) C(\bar{l} 1 \bar{l}+1; m_j-m_s, \pm 1 m_j-m_s \pm 1) \right. \right. \nonumber \\ &  & \left. \left. \times C(\bar{l}+1 1 \bar{l}; 0
 0 0) \right. \right. \nonumber \\
 &  & \left. \left. + \langle \bar{g}^\prime + \bar{f}^\prime \parallel j_L (k_\lambda r) \parallel
 F_{njl}^{(-)} \rangle C(\bar{l}-1 L \bar{l}^\prime ; m_j-m_s \pm 1 0 m_j-m_s
 \pm 1)  \right. \right. \nonumber \\
 &  & \left. \left. \times C(\bar{l}^\prime L \bar{l}-1; 0 0 0) C(\bar{l} 1 \bar{l}-1; m_j-m_s, \pm 1 m_j-m_s \pm 1) \right. \right. \nonumber \\
 &  & \left. \left. \times C(\bar{l}-1 1 \bar{l}; 0
 0 0) \right\} \right]
 \label{M-1-8C-2} \\
 M_2^{(\pm)} & = & \pm \hbar k_\lambda  \sqrt{2} i \sum_{L} (2L + 1)
 i^{-L} \langle \bar{g}^\prime + \bar{f}^\prime  \parallel
 j_L (k_\lambda r) \parallel \bar{g} + \bar{f} \rangle
 \nonumber \\
 &  & \times \left[\bar{\kappa}^\prime C(\bar{l}^\prime
 \mbox{$\frac{1}{2}$} j^\prime ; m_j \pm \mbox{$\frac{1}{2}$},
 \pm \mbox{$\frac{1}{2}$} m_j \pm 1) C(l \mbox{$\frac{1}{2}$} j;
 m_j \pm \mbox{$\frac{1}{2}$},\, \mp \mbox{$\frac{1}{2}$} m_j)
 \right. \nonumber \\
 &  & \left. \times C(l L \bar{l}^\prime; m_j \pm
 \mbox{$\frac{1}{2}$} 0 m_j \pm \mbox{$\frac{1}{2}$})
 C(\bar{l}^\prime L l ; 0 0 0) \right. \nonumber \\
 &  & \left. + \bar{\kappa}  C(l^\prime
 \mbox{$\frac{1}{2}$} j^\prime ; m_j \pm \mbox{$\frac{1}{2}$},
 \pm \mbox{$\frac{1}{2}$} m_j \pm 1) C(\bar{l} \mbox{$\frac{1}{2}$} j;
 m_j \pm \mbox{$\frac{1}{2}$}, \mp \mbox{$\frac{1}{2}$} m_j)
 \right. \nonumber \\
 &  & \left. \times C(\bar{l} L l^\prime; m_j \pm
 \mbox{$\frac{1}{2}$} 0 m_j \pm \mbox{$\frac{1}{2}$})
 C(l^\prime L \bar{l} ; 0 0 0) \right]
 \label{M-2-8C-2} \\
 M_3^{(\pm)} & = & \pm \hbar k_\lambda  \sqrt{2} \sum_{L} (2L + 1)
 i^{-L} \langle \bar{g}^\prime + \bar{f}^\prime \parallel
 j_L (k_\lambda r) \parallel \bar{g} + \bar{f} \rangle
 \nonumber \\
 &  & \times \left[ C(l^\prime
 \mbox{$\frac{1}{2}$} j^\prime ; m_j \pm \mbox{$\frac{1}{2}$},
 \pm \mbox{$\frac{1}{2}$} m_j \pm 1) C(l \mbox{$\frac{1}{2}$} j;
 m_j \pm \mbox{$\frac{1}{2}$}, \mp \mbox{$\frac{1}{2}$} m_j)
 \right. \nonumber \\
 &  & \left. \times C(l L l^\prime; m_j \pm
 \mbox{$\frac{1}{2}$} 0 m_j \pm \mbox{$\frac{1}{2}$})
 C(l^\prime L l ; 0 0 0) \right. \nonumber \\
 &  & \left. - \bar{\kappa}^\prime \bar{\kappa}  C(\bar{l}^\prime
 \mbox{$\frac{1}{2}$} j^\prime ; m_j \pm \mbox{$\frac{1}{2}$},
 \pm \mbox{$\frac{1}{2}$} m_j \pm 1) C(\bar{l} \mbox{$\frac{1}{2}$} j;
 m_j \pm \mbox{$\frac{1}{2}$}, \mp \mbox{$\frac{1}{2}$} m_j)
 \right. \nonumber \\
 &  & \left. \times C(\bar{l} L \bar{l}^\prime; m_j \pm
 \mbox{$\frac{1}{2}$} 0 m_j \pm \mbox{$\frac{1}{2}$})
 C(\bar{l}^\prime L \bar{l} ; 0 0 0) \right].
\label{M-3-8C-2}
\end{eqnarray}
Using (\ref{M-1-8C-2}), (\ref{M-2-8C-2}) and (\ref{M-3-8C-2}) in
the relation corresponding to (\ref{PT-formula}) :
\begin{equation}
P_T = \frac{2 e^2}{\hbar \mu^2 c^2} \sum_{m_{j^\prime}, m_j}
\sum_{\lambda} k_\lambda \left| M_1^\lambda + m_2^\lambda +
M_3^\lambda \right|^2 \label{PT-8C}
\end{equation}
gives the transition probabilities for the components of the
Balmer and Lyman $\alpha$-lines for various hydrogenic atoms
presented in the columns labelled 8-C in Tables~\ref{allow-1} -
\ref{forbid-92}.

%%%%%%%%%%%%%%%%%%%%%%%%%%%%%%%%%%%%%%%%%%%%%%
\section{Comparison of results and conclusion}
\label{result}
%%%%%%%%%%%%%%%%%%%%%%%%%%%%%%%%%%%%%%%%%%%%%%

Tables~\ref{allow-1} - \ref{allow-92} show the transition
probabilities (in $s^{-1}$) for the ``allowed'' components of the
Balmer and Lyman $\alpha$-lines for various hydrogenic atoms (Z=1,
18, 30, 54, 74 and 92). Tables~\ref{forbid-1} - \ref{forbid-92}
show the corresponding transition probabilities for the
``forbidden'' components, i.e. those components which are not
allowed in the usual dipole approximation~\cite{Friedrich90}. The
columns labelled S show the results given by the nonrelativistic
Schr\"odinger theory [eq. (\ref{PT-formula})]. The results for the
``allowed'' components for hydrogen agree with those of Condon and
Shortley~\cite{CS35}. The columns labelled D give the predictions
of  the Dirac theory [eq. (\ref{PT-diracfinal})]. These are in the
agreement with those of Pal'chikov~\cite{P98} and differ
considerably from the Schr\"odinger results for large Z. The
columns labelled 8-C show the results for the eight-component
theory [eq. (\ref{PT-8C})]. As for the Dirac theory, the
predictions of the 8-C theory for low Z are approximately the same
as the Schr\"odinger predictions. However, for larger values of Z,
it is seen that the results differ significantly from both the
Schr\"odinger and Dirac results. These results indicate for the
first time that the Dirac and 8-C theories are not identical in
all their predictions.

The calculated differences between the two relativistic formalisms
imply that in special circumstances it may be possible to
determine by observation which theory is valid. However, at the
present time, it is impossible to measure directly the transition
probabilities for the components of the Balmer and Lyman
$\alpha$-lines for high Z hydrogenic atoms since the lifetimes of
the excited states are too short so that the ``allowed''
transitions are prompt. On the other hand, the ``forbidden''
transitions are masked by ``allowed'' transitions, except for the
$3D_{5/2}$ to $2P_{1/2}$ or $2S_{1/2}$ transitions. These
``forbidden'' transitions may eventually be measurable for
intermediate Z values ($Z \simeq 50$) where the transition
probabilities are $\simeq 6 \times 10^{12} s^{-1}$. However for
$Z=50$, the differences between the Dirac and 8-C theories are
only $\simeq 5 \%$ for the dominant mode $3D_{5/2}$ to $2S_{1/2}$.

%%%%%%%%%%%%%%%%%%%%%%%%%%%%%%%
%REFERENCES

%%%%%%%%%%%%%%%%%%%%%%%%%%%%%%%%%%%%%%%%%%%%%%
%Tables
%%%%%%%%%%%%%%%%%%%%%%%%%%%%%%%%%%%%%%%%%%%%%%

\begin{table}
\begin{indented}
\item[]
\caption{Transition Probabilities (in $s^{-1}$) for the ``allowed''
components of the Balmer and Lyman $\alpha$-lines for $Z = 1$
hydrogenic atom for the Schr\"odinger (S), Dirac (D) and
eight-component (8-C) theories.} \label{allow-1}

\begin{tabular}{lccc}
 &   &   & \\
{\bf Transition} &  {\bf S} & {\bf D} & {\bf 8-C} \\ \hline $3
D_{5/2}$ - $2 P_{3/2}$ &  3.881 + 08 & 3.882 + 08  & 3.882 + 08
\\
$3 D_{3/2}$ - $2 P_{3/2}$ &  4.312 + 07 & 4.313 + 07  & 4.313 + 07
\\
$3 D_{3/2}$ - $2 P_{1/2}$ &  2.156 + 08 & 2.157 + 08  & 2.157 + 08
\\
$3 P_{3/2}$ - $2 S_{1/2}$ &  8.984 + 07 & 8.986 + 07  & 8.986 + 07
\\
$3 P_{1/2}$ - $2 S_{1/2}$ &  4.492 + 07 & 4.493 + 07  & 4.493 + 07
\\
$3 S_{1/2}$ - $2 P_{3/2}$ &  8.422 + 06 & 8.425 + 06  & 8.426 + 06
\\
$3 S_{1/2}$ - $2 P_{1/2}$ &  4.211 + 06 & 4.212 + 06  & 4.213 + 06
\\
$2 P_{3/2}$ - $1 S_{1/2}$ &  2.507 + 09 & 2.508 + 09  & 2.508 + 09
\\
$2 P_{1/2}$ - $1 S_{1/2}$ &  1.254 + 09 & 1.254 + 09  & 1.254 + 09
\\
\hline
\end{tabular}

\caption{Transition Probabilities (in $s^{-1}$) for the ``allowed''
components of the Balmer and Lyman $\alpha$-lines for $Z = 18$
hydrogenic atom for the Schr\"odinger (S), Dirac (D) and
eight-component (8-C) theories.} \label{allow-18}

\begin{tabular}{lccc}
 &   &   & \\
{\bf Transition} &  {\bf S} & {\bf D} & {\bf 8-C} \\ \hline $3
D_{5/2}$ - $2 P_{3/2}$ & 4.069 + 13  & 4.072 + 13  & 4.079 + 13
\\
$3 D_{3/2}$ - $2 P_{3/2}$ & 4.521 + 12  & 4.523 + 12  & 4.501 + 12
\\
$3 D_{3/2}$ - $2 P_{1/2}$ & 2.261 + 13  & 2.280 + 13  & 2.279 + 13
\\
$3 P_{3/2}$ - $2 S_{1/2}$ & 9.419 + 12  & 9.392 + 12  & 9.401 + 12
\\
$3 P_{1/2}$ - $2 S_{1/2}$ & 4.710 + 12  & 4.772 + 12  & 4.803 + 12
\\
$3 S_{1/2}$ - $2 P_{3/2}$ & 8.830 + 11  & 9.263 + 11  & 9.308 + 11
\\
$3 S_{1/2}$ - $2 P_{1/2}$ & 4.415 + 11  & 4.470 + 11  & 4.639 + 11
\\
$2 P_{3/2}$ - $1 S_{1/2}$ & 2.621 + 14  & 2.622 + 14  & 2.645 + 14
\\
$2 P_{1/2}$ - $1 S_{1/2}$ & 1.310 + 14  & 1.319 + 14  & 1.301 + 14
\\
\hline
\end{tabular}

\caption{Transition Probabilities (in $s^{-1}$) for the ``allowed''
components of the Balmer and Lyman $\alpha$-lines for $Z = 30$
hydrogenic atom for the Schr\"odinger (S), Dirac (D) and
eight-component (8-C) theories.} \label{allow-30}

\begin{tabular}{lccc}
 &   &   & \\
{\bf Transition} &  {\bf S} & {\bf D} & {\bf 8-C} \\ \hline $3
D_{5/2}$ - $2 P_{3/2}$ & 3.133 + 14  & 3.138 + 14  & 3.153 + 14
\\
$3 D_{3/2}$ - $2 P_{3/2}$ & 3.481 + 13  & 3.483 + 13  & 3.435 + 13
\\
$3 D_{3/2}$ - $2 P_{1/2}$ & 1.740 + 14  & 1.782 + 14  & 1.779 + 14
\\
$3 P_{3/2}$ - $2 S_{1/2}$ & 7.252 + 13  & 7.185 + 13  & 7.203 + 13
\\
$3 P_{1/2}$ - $2 S_{1/2}$ & 3.626 + 13  & 3.760 + 13  & 3.831 + 13
\\
$3 S_{1/2}$ - $2 P_{3/2}$ & 6.798 + 12  & 7.748 + 12  & 7.853 + 12
\\
$3 S_{1/2}$ - $2 P_{1/2}$ & 3.399 + 12  & 3.518 + 12  & 3.902 + 12
\\
$2 P_{3/2}$ - $1 S_{1/2}$ & 2.007 + 15  & 2.009 + 15  & 2.058 + 15
\\
$2 P_{1/2}$ - $1 S_{1/2}$ & 1.003 + 15  & 1.021 + 15  & 9.820 + 14
\\
\hline
\end{tabular}
\end{indented}
\end{table}

\begin{table}
\begin{indented}
\item[]
\caption{Transition Probabilities (in $s^{-1}$) for the ``allowed''
components of the Balmer and Lyman $\alpha$-lines for $Z = 54$
hydrogenic atom for the Schr\"odinger (S), Dirac (D) and
eight-component (8-C) theories.} \label{allow-54}

\begin{tabular}{lccc}
 &   &   & \\
{\bf Transition} &  {\bf S} & {\bf D} & {\bf 8-C} \\ \hline $3
D_{5/2}$ - $2 P_{3/2}$ & 3.263 + 15  & 3.278 + 15 & 3.329 + 15
\\
$3 D_{3/2}$ - $2 P_{3/2}$ & 3.626 + 14  & 3.629 + 14 & 3.471 + 14
\\
$3 D_{3/2}$ - $2 P_{1/2}$ & 1.813 + 15  & 1.957 + 15 & 1.948 + 15
\\
$3 P_{3/2}$ - $2 S_{1/2}$ & 7.554 + 14  & 7.248 + 14 & 7.295 + 14
\\
$3 P_{1/2}$ - $2 S_{1/2}$ & 3.777 + 14  & 4.268 + 14 & 4.562 + 14
\\
$3 S_{1/2}$ - $2 P_{3/2}$ & 7.081 + 13  & 1.062 + 14 & 1.110 + 14
\\
$3 S_{1/2}$ - $2 P_{1/2}$ & 3.541 + 13  & 3.970 + 13 & 5.614 + 13
\\
$2 P_{3/2}$ - $1 S_{1/2}$ & 2.051 + 16  & 2.053 + 16 & 2.225 + 16
\\
$2 P_{1/2}$ - $1 S_{1/2}$ & 1.026 + 16  & 1.086 + 16 & 9.523 + 15
\\
\hline
\end{tabular}

\caption{Transition Probabilities (in $s^{-1}$) for the ``allowed''
components of the Balmer and Lyman $\alpha$-lines for $Z = 74$
hydrogenic atom for the Schr\"odinger (S), Dirac (D) and
eight-component (8-C) theories.} \label{allow-74}

\begin{tabular}{lccc}
 &   &   & \\
{\bf Transition} &  {\bf S} & {\bf D} & {\bf 8-C} \\ \hline $3
D_{5/2}$ - $2 P_{3/2}$ & 1.140 + 16  & 1.149 + 16  & 1.183 + 16
\\
$3 D_{3/2}$ - $2 P_{3/2}$ & 1.266 + 15  & 1.267 + 15  & 1.165 + 15
\\
$3 D_{3/2}$ - $2 P_{1/2}$ & 6.331 + 15  & 7.299 + 15  & 7.262 + 15
\\
$3 P_{3/2}$ - $2 S_{1/2}$ & 2.638 + 15  & 2.360 + 15  & 2.372 + 15
\\
$3 P_{1/2}$ - $2 S_{1/2}$ & 1.319 + 15  & 1.680 + 15  & 1.943 + 15
\\
$3 S_{1/2}$ - $2 P_{3/2}$ & 2.473 + 14  & 5.085 + 14  & 5.557 + 14
\\
$3 S_{1/2}$ - $2 P_{1/2}$ & 1.236 + 14  & 1.548 + 14  & 3.066 + 14
\\
$2 P_{3/2}$ - $1 S_{1/2}$ & 6.994 + 16  & 6.972 + 16  & 8.153 + 16
\\
$2 P_{1/2}$ - $1 S_{1/2}$ & 3.497 + 16  & 3.891 + 16  & 3.004 + 16
\\
\hline
\end{tabular}

\caption{Transition Probabilities (in $s^{-1}$) for the ``allowed''
components of the Balmer and Lyman $\alpha$-lines for $Z = 92$
hydrogenic atom for the Schr\"odinger (S), Dirac (D) and
eight-component (8-C) theories.} \label{allow-92}

\begin{tabular}{lccc}
 &   &   & \\
{\bf Transition} &  {\bf S} & {\bf D} & {\bf 8-C} \\ \hline $3
D_{5/2}$ - $2 P_{3/2}$ & 2.691 + 16  & 2.725 + 16  & 2.852 + 16
\\
$3 D_{3/2}$ - $2 P_{3/2}$ & 2.991 + 15  & 2.992 + 15  & 2.626 + 15
\\
$3 D_{3/2}$ - $2 P_{1/2}$ & 1.495 + 16  & 1.856 + 16  & 1.857 + 16
\\
$3 P_{3/2}$ - $2 S_{1/2}$ & 6.231 + 15  & 4.796 + 15  & 4.721 + 15
\\
$3 P_{1/2}$ - $2 S_{1/2}$ & 3.115 + 15  & 4.658 + 15  & 6.134 + 15
\\
$3 S_{1/2}$ - $2 P_{3/2}$ & 5.839 + 14  & 1.666 + 15  & 1.939 + 15
\\
$3 S_{1/2}$ - $2 P_{1/2}$ & 2.919 + 14  & 4.228 + 14  & 1.316 + 15
\\
$2 P_{3/2}$ - $1 S_{1/2}$ & 1.607 + 17  & 1.580 + 17  & 2.035 + 17
\\
$2 P_{1/2}$ - $1 S_{1/2}$ & 8.036 + 16  & 9.454 + 16  & 6.138 + 16
\\
\hline
\end{tabular}
\end{indented}
\end{table}

\begin{table}
\begin{indented}
\item[]
\caption{Transition Probabilities (in $s^{-1}$) for the
``forbidden'' components of the Balmer and Lyman $\alpha$-lines for
$Z = 1$ hydrogenic atom for the Schr\"odinger (S), Dirac (D) and
eight-component (8-C) theories.} \label{forbid-1}

\begin{tabular}{lccc}
 &   &   & \\
{\bf Transition} &  {\bf S} & {\bf D} & {\bf 8-C} \\ \hline $3
D_{5/2}$ - $2 P_{1/2}$ & 1.344 - 04  & 3.115 - 04  & 5.309 - 04
\\
$3 D_{5/2}$ - $2 S_{1/2}$ & 3.062 + 02  & 3.063 + 02  & 3.063 + 02
\\
$3 D_{3/2}$ - $2 S_{1/2}$ & 2.041 + 02  & 2.042 + 02  & 2.042 + 02
\\
$3 P_{3/2}$ - $2 P_{3/2}$ & 4.784 + 01  & 4.785 + 01  & 4.785 + 01
\\
$3 P_{3/2}$ - $2 P_{1/2}$ & 4.784 + 01  & 4.785 + 01  & 4.785 + 01
\\
$3 P_{1/2}$ - $2 P_{3/2}$ & 4.784 + 01  & 4.785 + 01  & 4.785 + 01
\\
$3 P_{1/2}$ - $2 P_{1/2}$ & 1.090 - 10  & 9.816 - 10  & 2.727 - 09
\\
$3 S_{1/2}$ - $2 S_{1/2}$ & 0.000 + 00  & 3.756 - 09  & 3.756 - 09
\\
$2 S_{1/2}$ - $1 S_{1/2}$ & 0.000 + 00  & 4.993 - 06  & 4.993 - 06
\\
\hline
\end{tabular}
%\newpage

\caption{Transition Probabilities (in $s^{-1}$) for the
``forbidden'' components of the Balmer and Lyman $\alpha$-lines for
$Z = 18$ hydrogenic atom for the Schr\"odinger (S), Dirac (D) and
eight-component (8-C) theories.} \label{forbid-18}

\begin{tabular}{lccc}
 &   &   & \\
{\bf Transition} &  {\bf S} & {\bf D} & {\bf 8-C} \\ \hline $3
D_{5/2}$ - $2 P_{1/2}$ & 1.480 + 06  & 3.524 + 06  & 6.051 + 06
\\
$3 D_{5/2}$ - $2 S_{1/2}$ & 1.040 + 10  & 1.065 + 10  & 1.072 + 10
\\
$3 D_{3/2}$ - $2 S_{1/2}$ & 6.934 + 09  & 7.081 + 09  & 7.114 + 09
\\
$3 P_{3/2}$ - $2 P_{3/2}$ & 1.626 + 09  & 1.626 + 09  & 1.627 + 09
\\
$3 P_{3/2}$ - $2 P_{1/2}$ & 1.626 + 09  & 1.640 + 09  & 1.660 + 09
\\
$3 P_{1/2}$ - $2 P_{3/2}$ & 1.626 + 09  & 1.632 + 09  & 1.635 + 09
\\
$3 P_{1/2}$ - $2 P_{1/2}$ & 3.889 + 02  & 3.603 + 03  & 1.019 + 04
\\
$3 S_{1/2}$ - $2 S_{1/2}$ & 0.000 + 00  & 1.379 + 04  & 1.396 + 04
\\
$2 S_{1/2}$ - $1 S_{1/2}$ & 0.000 + 00  & 1.816 + 07  & 1.838 + 07
\\
\hline
\end{tabular}
%\newpage

\caption{Transition Probabilities (in $s^{-1}$) for the
``forbidden'' components of the Balmer and Lyman $\alpha$-lines for
$Z = 30$ hydrogenic atom for the Schr\"odinger (S), Dirac (D) and
eight-component (8-C) theories.} \label{forbid-30}

\begin{tabular}{lccc}
 &   &   & \\
{\bf Transition} &  {\bf S} & {\bf D} & {\bf 8-C} \\ \hline $3
D_{5/2}$ - $2 P_{1/2}$ & 8.798 + 07  & 2.199 + 08  & 3.829 + 08
\\
$3 D_{5/2}$ - $2 S_{1/2}$ & 2.224 + 11  & 2.378 + 11  & 2.422 + 11
\\
$3 D_{3/2}$ - $2 S_{1/2}$ & 1.483 + 11  & 1.572 + 11  & 1.593 + 11
\\
$3 P_{3/2}$ - $2 P_{3/2}$ & 3.478 + 10  & 3.480 + 10  & 3.486 + 10
\\
$3 P_{3/2}$ - $2 P_{1/2}$ & 3.478 + 10  & 3.561 + 10  & 3.689 + 10
\\
$3 P_{1/2}$ - $2 P_{3/2}$ & 3.478 + 10  & 3.511 + 10  & 3.535 + 10
\\
$3 P_{1/2}$ - $2 P_{1/2}$ & 6.421 + 04  & 6.268 + 05  & 1.833 + 06
\\
$3 S_{1/2}$ - $2 S_{1/2}$ & 0.000 + 00  & 2.397 + 06  & 2.485 + 06
\\
$2 S_{1/2}$ - $1 S_{1/2}$ & 0.000 + 00  & 3.105 + 09  & 3.217 + 09
\\
\hline
\end{tabular}
\end{indented}
\end{table}

\begin{table}
\begin{indented}
\item[]
\caption{Transition Probabilities (in $s^{-1}$) for the
``forbidden'' components of the Balmer and Lyman $\alpha$-lines for
$Z = 54$ hydrogenic atom for the Schr\"odinger (S), Dirac (D) and
eight-component (8-C) theories.} \label{forbid-54}

\begin{tabular}{lccc}
 &   &   & \\
{\bf Transition} &  {\bf S} & {\bf D} & {\bf 8-C} \\ \hline $3
D_{5/2}$ - $2 P_{1/2}$ & 9.638 + 09  & 2.868 + 10  & 5.265 + 10
\\
$3 D_{5/2}$ - $2 S_{1/2}$ & 7.506 + 12  & 9.335 + 12  & 9.945 + 12
\\
$3 D_{3/2}$ - $2 S_{1/2}$ & 5.004 + 12  & 6.079 + 12  & 6.375 + 12
\\
$3 P_{3/2}$ - $2 P_{3/2}$ & 1.176 + 12  & 1.180 + 12  & 1.185 + 12
\\
$3 P_{3/2}$ - $2 P_{1/2}$ & 1.176 + 12  & 1.268 + 12  & 1.441 + 12
\\
$3 P_{1/2}$ - $2 P_{3/2}$ & 1.176 + 12  & 1.205 + 12  & 1.256 + 12
\\
$3 P_{1/2}$ - $2 P_{1/2}$ & 2.279 + 07  & 2.696 + 08  & 8.938 + 08
\\
$3 S_{1/2}$ - $2 S_{1/2}$ & 0.000 + 00  & 1.030 + 09  & 1.167 + 09
\\
$2 S_{1/2}$ - $1 S_{1/2}$ & 0.000 + 00  & 1.255 + 12  & 1.418 + 12
\\
\hline
\end{tabular}

%\newpage

\caption{Transition Probabilities (in $s^{-1}$) for the
``forbidden'' components of the Balmer and Lyman $\alpha$-lines for
$Z = 74$ hydrogenic atom for the Schr\"odinger (S), Dirac (D) and
eight-component (8-C) theories.} \label{forbid-74}

\begin{tabular}{lccc}
 &   &   & \\
{\bf Transition} &  {\bf S} & {\bf D} & {\bf 8-C} \\ \hline $3
D_{5/2}$ - $2 P_{1/2}$ & 1.190 + 11  & 4.452 + 11  & 8.861 + 11
\\
$3 D_{5/2}$ - $2 S_{1/2}$ & 4.922 + 13  & 7.462 + 13  & 8.489 + 13
\\
$3 D_{3/2}$ - $2 S_{1/2}$ & 3.281 + 13  & 4.795 + 13  & 5.320 + 13
\\
$3 P_{3/2}$ - $2 P_{3/2}$ & 7.731 + 12  & 7.811 + 12  & 7.837 + 12
\\
$3 P_{3/2}$ - $2 P_{1/2}$ & 7.730 + 12  & 8.871 + 12  & 1.178 + 13
\\
$3 P_{1/2}$ - $2 P_{3/2}$ & 7.730 + 12  & 7.957 + 12  & 9.067 + 12
\\
$3 P_{1/2}$ - $2 P_{1/2}$ & 5.283 + 08  & 8.181 + 09  & 3.264 + 10
\\
$3 S_{1/2}$ - $2 S_{1/2}$ & 0.000 + 00  & 3.123 + 10  & 4.040 + 10
\\
$2 S_{1/2}$ - $1 S_{1/2}$ & 0.000 + 00  & 3.489 + 13  & 4.477 + 13
\\
\hline
\end{tabular}
%\newpage

\caption{Transition Probabilities (in $s^{-1}$) for the
``forbidden'' components of the Balmer and Lyman $\alpha$-lines for
$Z = 92$ hydrogenic atom for the Schr\"odinger (S), Dirac (D) and
eight-component (8-C) theories.} \label{forbid-92}

\begin{tabular}{lccc}
 &   &   & \\
{\bf Transition} &  {\bf S} & {\bf D} & {\bf 8-C} \\ \hline $3
D_{5/2}$ - $2 P_{1/2}$ & 6.730 + 11  & 3.338 + 12  & 7.511 + 12
\\
$3 D_{5/2}$ - $2 S_{1/2}$ & 1.797 + 14  & 3.469 + 14  & 4.337 + 14
\\
$3 D_{3/2}$ - $2 S_{1/2}$ & 1.198 + 14  & 2.219 + 14  & 2.694 + 14
\\
$3 P_{3/2}$ - $2 P_{3/2}$ & 2.830 + 13  & 2.900 + 13  & 2.892 + 13
\\
$3 P_{3/2}$ - $2 P_{1/2}$ & 2.829 + 13  & 3.478 + 13  & 6.018 + 13
\\
$3 P_{1/2}$ - $2 P_{3/2}$ & 2.829 + 13  & 2.848 + 13  & 3.909 + 13
\\
$3 P_{1/2}$ - $2 P_{1/2}$ & 4.619 + 09  & 1.029 + 11  & 5.367 + 11
\\
$3 S_{1/2}$ - $2 S_{1/2}$ & 0.000 + 00  & 3.930 + 11  & 6.179 + 11
\\
$2 S_{1/2}$ - $1 S_{1/2}$ & 0.000 + 00  & 3.894 + 14  & 6.005 + 14
\\
\hline
\end{tabular}
\end{indented}
\end{table}

\end{document}